\documentclass[aps,prd,preprintnumbers,superscriptaddress,nofootinbib]{revtex4-1}
\usepackage{amsmath}
\usepackage{amssymb}
\usepackage{amsthm}
\usepackage{mathtools}
\usepackage{mathrsfs}
\usepackage{bm}
\usepackage{slashed} 
\usepackage{graphicx}
\usepackage{multirow}
\usepackage{tikz}
\usepackage[caption=false]{subfig}
\usepackage{relsize}	
\usepackage{array}
\usepackage{float}
\usepackage{color}
\usepackage{xcolor}
\usepackage{soul}
\usepackage{verbatim} 

\allowdisplaybreaks

\newcommand{\seq}{\begin{subequations}}
\newcommand{\sen}{\end{subequations}}
\newcommand{\be}{\begin{eqnarray}}
\newcommand{\ee}{\end{eqnarray}}
\newcommand{\eq}{\begin{eqnarray}}
\newcommand{\en}{\end{eqnarray}}

\newcommand{\bfk}{{\bf k}_{\perp}}

\newcommand{\bfP}{{\bf P}_{\perp}} 
 
\newcommand{\bfp}{{\bf p}_{\perp}}

\newcommand{\nn}{\nonumber}

\usepackage{booktabs}
\AtBeginDocument{
\heavyrulewidth=.08em
\lightrulewidth=.05em
\cmidrulewidth=.03em
\belowrulesep=.65ex
\belowbottomsep=0pt
\aboverulesep=.4ex
\abovetopsep=0pt
\cmidrulesep=\doublerulesep
\cmidrulekern=.5em
\defaultaddspace=.5em
}

\begin{document}

\title{ Gravitational form factors and mechanical properties of proton in a light-front quark-diquark model }

\author{Dipankar Chakrabarti}
\email{dipankar@iitk.ac.in} \affiliation{ Department of Physics,
Indian Institute of Technology Kanpur, Kanpur 208016, India}

\author{Chandan Mondal}
\email{mondal@impcas.ac.cn} \affiliation{Institute of Modern Physics, Chinese Academy of Sciences, Lanzhou 730000, China}
\affiliation{School of Nuclear Science and Technology, University of Chinese Academy of Sciences, Beijing 100049, China}
\affiliation{CAS Key Laboratory of High Precision Nuclear Spectroscopy, Institute of Modern Physics, Chinese Academy of Sciences, Lanzhou 730000, China}

\author{Asmita Mukherjee}
\email{asmita@phy.iitb.ac.in} \affiliation{ Department of Physics,
Indian Institute of Technology Bombay,Powai, Mumbai 400076,
India}

\author{Sreeraj Nair}
\email{sreeraj@impcas.ac.cn} \affiliation{Institute of Modern Physics, Chinese Academy of Sciences, Lanzhou 730000, China}
\affiliation{School of Nuclear Science and Technology, University of Chinese Academy of Sciences, Beijing 100049, China}
\affiliation{CAS Key Laboratory of High Precision Nuclear Spectroscopy, Institute of Modern Physics, Chinese Academy of Sciences, Lanzhou 730000, China}

\author{Xingbo Zhao}
\email{xbzhao@impcas.ac.cn} \affiliation{Institute of Modern Physics, Chinese Academy of Sciences, Lanzhou 730000, China}
\affiliation{School of Nuclear Science and Technology, University of Chinese Academy of Sciences, Beijing 100049, China}
\affiliation{CAS Key Laboratory of High Precision Nuclear Spectroscopy, Institute of Modern Physics, Chinese Academy of Sciences, Lanzhou 730000, China}

\date{\today}

\begin{abstract}
We obtain the gravitational
form factors (GFFs) and investigate their applications for the description of the mechanical properties, i.e., the distributions of pressures, shear forces inside proton, and the mechanical radius, in a light-front quark-diquark model constructed by the soft-wall AdS/QCD. The GFFs, $A(Q^2)$ and $B(Q^2)$ are found to be consistent with the lattice QCD, while the qualitative behavior of the $D$-term form factor is in agreement with the extracted data from the deeply virtual Compton scattering (DVCS) experiments at JLab, the lattice QCD, and the predictions of different phenomenological models. The pressure and shear force distributions are also consistent with the results of different models.

\end{abstract}

\maketitle
\section{Introduction}
The mechanical properties of the nucleon, namely how the mass, spin and
pressure are distributed
among the quarks and gluons inside the nucleon is a topic of intense
interest in recent
days~\cite{Lorce:2018egm,Burkert:2018bqq,Mondal:2015fok,Anikin:2019kwi,Polyakov:2018zvc,Goeke:2007fp,Azizi:2019ytx,Jung:2014jja,Cebulla:2007ei,Kim:2012ts,Shanahan:2018nnv,Shanahan:2018pib,Hagler:2003jd,Gockeler:2003jfa,Gockeler:2003jfa,Pasquini:2014vua,Polyakov:2018exb,Polyakov:2002yz,Mamo:2019mka,Abidin:2009hr,Varma:2020crx,Neubelt:2019sou,Liuti:2018ccr,Kumericki:2015lhb,Tanaka:2018nae,Hatta:2018sqd,Ji:1997gm}. These information are related to the gravitomagnetic form factors,
which are
expressed as the matrix elements of the energy-momentum tensor in the
proton state. The components
of the energy-momentum tensor give how matter couples to the
gravitational field.
Thus these form factors can be obtained by direct measurement of the
interaction of the proton
with a strong gravitational field for example a neutron star. An indirect
way to obtain information
on them is from hard exclusive processes for example deeply virtual
Compton scattering (DVCS)
that is sensitive to the gravitational form factors (GFFs) through generalized
parton distributions (GPDs)~\cite{Burkert:2018bqq}.
The GFFs are functions of $t=-Q^2$, which is the
squared momentum transfer
 from the initial to final proton in DVCS experiment. The GFFs $A(Q^2)$ and
$B(Q^2)$ are related to the
 mass  and spin of the proton. The Ji's sum~\cite{Ji:1996ek} rule relates the second Mellin's
moment of the GPDs
$H$ and $E$ to the quark contribution to the angular momentum $J$. First
experimental results
relevant for the extraction of GPDs were provided by HERA~\cite{Adloff:1999kg,Adloff:2001cn,Breitweg:1998nh,Chekanov:2003ya,Chekanov:2003ya}, HERMES~\cite{Airapetian:2001yk}, COMPASS~\cite{dHose:2004usi} and
JLab~\cite{Stepanyan:2001sm,Diehl:2020uja}. These are also being
investigated at JLab 12 GeV upgrade and COMPASS at CERN, and will be
accessed at the future
electron-ion collider (EIC)~\cite{Accardi:2012qut}.

The GFFs $A$ and $B$ are related to the generators of the Poincare
group, which gives constraints on them at zero momentum transfer, that
helps in the extraction
of these form factors from the experimental data. In contrast the GFF $C(Q^2)$
(also called the D-term) is unconstrained at zero momentum transfer. This
form factor is related to the
internal properties of the nucleon like the pressure and stress
distribution~\cite{Polyakov:2018zvc,Lorce:2018egm}. This form factor contributes
to the DVCS process when the skewness $\xi$ is non-zero, or when there is
non-zero momentum
transfer in the longitudinal direction. $C$ form factor has been
calculated in several models
in the literature. It depends on the correlations between the quarks and
gluons in the nucleon.
The GFFs of the nucleon have been investigated in the framework of lattice QCD 
\cite{Hagler:2003jd,Gockeler:2003jfa,Bratt:2010jn, Hagler:2007xi,Brommel:2007sb,Deka:2013zha}, 
chiral perturbation theory ($\chi$PT) 
\cite{Chen:2001pva,Belitsky:2002jp,Dorati:2007bk}, 
 Skyrme model \cite{Cebulla:2007ei,Kim:2012ts}, chiral quark soliton model ($\chi$QSM) \cite{ Schweitzer:2002nm, Wakamatsu:2006dy, Goeke:2007fq,Goeke:2007fp,Wakamatsu:2007uc}, light-cone QCD sum rules at leading order~(LCSR-LO)~\cite{Anikin:2019kwi}, dispersion relation (DR)~\cite{Pasquini:2014vua}, instanton picture (IP)~\cite{Polyakov:2018exb}, and instant and front form (IFF) \cite{Lorce:2018egm}, while the asymptotic behavior of the GFFs has been discussed in Refs.~\cite{Hatta:2018sqd,Tanaka:2018nae}. 
The $D-$term of a free spinless boson is $-1$ whereas for a free fermion
it is zero~\cite{Polyakov:2018zvc}. In fact
for an interacting system, stability requires that the form factor $C$ is negative.
This form factor has been
calculated in MIT bag model~\cite{Neubelt:2019sou}. 
It has also been
extracted from JLab data~\cite{Burkert:2018bqq}.
The bag model underestimates the data while the Skryme model overestimates
it. On the other hand
predictions from dispersion relation and $\chi$QSM are
more close to the data.
Renewed interest in the form factor $C (Q^2)$ was generated after a recent result from
JLab showed that the
pressure distribution is repulsive at the center of the nucleon and
confining towards the
outer region~\cite{Burkert:2018bqq,Shanahan:2018nnv}. At the center, it exceeds the pressure
estimated for the most dense object in the universe that are the neutron
stars. The anisotropy
of highly dense nuclear matter has been investigated in the literature.
This depends on the
interaction of nuclear matter inside neutron stars and cannot be explained
by the equation
of state~(see Ref.~\cite{Lorce:2018egm} and the references therein). Thus the study of the anisotropic pressure distribution
inside the nucleon
through the GPDs provides an indirect way to investigate such properties
in highly dense
astronomical objects~\cite{Liuti:2018ccr}. Initial theoretical studies of the pressure
distribution have been
formulated in the Breit frame. The distributions defined in this frame are
subject to
relativistic corrections. In \cite{Lorce:2018egm}, the pressure and energy distributions
inside a nucleon
are defined in different frames, including the Breit frame and infinite
momentum frame
or light-front formalism. The latter has the advantage that because of the
fact that
transverse boosts are Galilean in light-front framework, one can obtain a
relativistic
description of the form factors in terms of the light-front wave functions (LFWFs). In Ref.~\cite{Lorce:2018egm},
the energy
and pressure distributions are investigated by assuming a simple multipole
model for
 the GFFs. In this work we use a quark-diquark model based on
AdS/QCD to calculate
 the GFFs as well as the pressure distributions and compare with other model results in the literature.  The light-front wave functions in this model are constructed
from the two particle effective wave functions obtained in soft-wall of AdS/QCD~\cite{Brodsky:2007hb,BT}.
This model is consistent with Drell–Yan–West relation~\cite{Gutsche:2013zia} and has been successfully applied to
describe many interesting properties of nucleon e.g., electromagnetic form factor, PDFs, GPDs, TMDs, Wigner distributions, transverse densities etc.~\cite{Gutsche:2013zia,Chakrabarti:2016yuw, Chakrabarti:2015ama, Chakrabarti:2015lba, Mondal:2015uha,Gutsche:2016gcd,Mondal:2017wbf,Mondal:2016xsm,Maji:2015vsa}.

The paper is organized as follows. In section~\ref{model}, we give brief
introductions to the nucleon LFWFs of the quark–diquark model. The GFFs of proton have been evaluated in this model and discussed in section~\ref{gffs}. We study the mechanical properties of proton, e.g., the pressures, energy density distributions, shear forces, and the mechanical radius in section~\ref{properties}. Summary is
given in section~\ref{summary}.

\section{Light-front quark-diquark model} \label{model}

Here we adopt the generic ansatz for the light-front quark-diquark model for the nucleons \cite{Gutsche:2013zia} where the light-front wave functions
are modeled from the solution of soft-wall AdS/QCD. In this model, one contemplates the three valence quarks of the nucleons as an effective system
composed of a  quark (fermion) and a composite state of diquark (boson), where the  spin of the diquark is assumed to be zero (scalar) only.
Then the 2-particle Fock-state expansion for proton spin components, $J^z = \pm \frac{1}{2}$ in a frame where the transverse momentum of proton vanishes i.e. $P \equiv \big(P^+,\textbf{0}_\perp,\frac{M^2}{P^+}\big)$, 
is written as
 \be\label{state}
  |P;\uparrow(\downarrow)\rangle 
& =& \sum_q \int \frac{dx~ d^2\textbf{k}_{\perp}}{2(2\pi)^3\sqrt{x(1-x)}}\bigg[ \psi^{\uparrow(\downarrow)}_{+q}(x,\textbf{k}_{\perp})|+\frac{1}{2},0; xP^+,\textbf{k}_{\perp}\rangle + \psi^{\uparrow(\downarrow)}_{-q}(x,\textbf{k}_{\perp})|-\frac{1}{2},0; xP^+,\textbf{k}_{\perp}\rangle\bigg].
  \ee
However, for nonzero transverse momentum of proton, i.e., $\bfP\ne0$, the physical transverse momenta of quark and diquark are $\bfp^q=x\bfP+\bfk$ and $\bfp^D=(1-x)\bfP-\bfk$, respectively, where $\bfk$ represents the relative transverse momentum of the constituents. $\psi_{\lambda_q}^{\lambda_N}(x,\bfk)$ are the light-front wave functions with nucleon helicities $\lambda_N=\uparrow(\downarrow)$ and for quark $\lambda_q=\pm$; plus and minus
correspond to $+\frac{1}{2}$ and $-\frac{1}{2}$,  respectively. The light-front wave functions at an initial scale $\mu^2_0=0.32$ GeV$^2$ are given by  
\be\label{WF}
\psi_{+q}^\uparrow(x,\bfk) &=&  \varphi_q^{(1)}(x,\bfk) \,,\nonumber\\ 
\quad
\psi_{-q}^\uparrow(x,\bfk) &=& -\frac{k^1 + ik^2}{xM}   \, \varphi_q^{(2)}(x,\bfk) \,, \nonumber\\
\psi_{+q}^\downarrow(x,\bfk) &=& \frac{k^1 - ik^2}{xM}  \, \varphi_q^{(2)}(x,\bfk)\,. \\
\psi_{-q}^\downarrow(x,\bfk) &=& \varphi_q^{(1)}(x,\bfk),\nonumber
\ee
where, the wave functions $\varphi_q^{(i=1,2)}(x,\bfk)$ are the modified form of the soft-wall AdS/QCD prediction constructed by introducing the parameters $a_q^{(i)}$ and $b_q^{(i)}$ for quark $q$~\cite{BT,Gutsche:2013zia},
\be\label{wf2}
\varphi_q^{(i)}(x,\bfk)&=&N_q^{(i)}\frac{4\pi}{\kappa}\sqrt{\frac{\log(1/x)}{1-x}}x^{a_q^{(i)}}
(1-x)^{b_q^{(i)}}\exp\bigg[-\frac{\bfk^2}{2\kappa^2}\frac{\log(1/x)}{(1-x)^2}\bigg].
\ee
$\varphi_q^{(i)}(x,\bfk)$ reduces to the original AdS/QCD solution when $a_q^{(i)}=b_q^{(i)}=0$ 
\cite{BT}. It should be mentioned here that the modification of the soft-wall AdS/QCD prediction in Eq.~(\ref{wf2}) is not unique, while a generic reparametrization function $w(x)$, which unifies the description of polarized and unpolarized quark distributions in the proton, has been introduced in Refs.~\cite{deTeramond:2018ecg,Liu:2019vsn}. 
 In this work, we take the AdS/QCD scale parameter $\kappa =0.4$ GeV, obtained by
fitting the nucleon form factors in the soft-wall model of AdS/QCD \cite{CM1,CM2}. The quarks are assumed to be massless and the
parameters $a^{(i)}_q$ and $b^{(i)}_q$ with the constants $N^{(i)}_q$ are obtained by
fitting the electromagnetic properties of the nucleons: $F_1^q(0)=n_q$ and $F_2^q(0)=\kappa_q$
where $n_u=2$ and $n_d=1$, the number of valence $u$ and $d$ quarks in proton and
the anomalous magnetic moments for the $u$ and $d$ quarks are $\kappa_u=1.673$ and
$\kappa_d=-2.033$ \cite{Chakrabarti:2015ama,Mondal:2017wbf}.  Since no isospin or flavor symmetry is imposed, the parameters for $u$ and $d$ quarks in the model  are different. The parameters are given by  $a^{(1)}_u  = 0.020,~  a^{(1)}_d= 0.10,~
b^{(1)}_u = 0.022,~b^{(1)}_d=0.38,~
a^{(2)}_u=  1.033,~ a^{(2)}_d=  1.087,~
b^{(2)}_u= -0.15, ~b^{(2)}_d= -0.20,
N^{(1)}_u = 2.055,~ N^{(1)}_d = 1.7618,
N^{(2)}_u= 1.322, N^{(2)}_d = -2.4827$ and the quarks are assumed to be massless.
We estimate a $2\%$ uncertainty
in the model parameters. The model inspired by
soft-wall AdS/QCD has been extensively used to investigate and reproduce
many interesting properties of the nucleons 
\cite{Gutsche:2013zia,Chakrabarti:2016yuw, Chakrabarti:2015ama, Chakrabarti:2015lba, Mondal:2015uha,Gutsche:2016gcd,Mondal:2017wbf,Mondal:2016xsm,Maji:2015vsa}.

\section{Gravitational Form factors}\label{gffs}
 The matrix elements of local operators like electromagnetic current and energy momentum tensor   have exact representation in light-front Fock state wave functions of bound states such as hadrons. The gravitational form factors (GFFs) are related to the matrix elements of the energy-momentum tensor, $T^{\mu \nu}$, while the second moment of the GPDs also provides the GFFs. For a spin $1/2$ target, the standard parametrization of $T^{\mu \nu}$ involving the GFFs reads \cite{hari,ji12}
\be
\langle P', S'|T^{\mu \nu}_i(0)|P, S\rangle &=&\bar{U}(P', S')\bigg[-B_i(q^2)\frac{\bar{P}^\mu\bar{P^\nu}}{M} +(A_i(q^2)+B_i(q^2))\frac{1}{2}(\gamma^{\mu}\bar{P}^{\nu}+\gamma^{\nu}\bar{P}^{\mu}) \nonumber\\
&+& C_i(q^2)\frac{q^{\mu}q^{\nu}-q^2 g^{\mu\nu}}{M}+\bar{C}_i(q^2)M g^{\mu\nu}\bigg]U(P,S),\label{tensor}
\ee
where, $\bar{P}^\mu=\frac{1}{2}(P'+P)^\mu$, $q^\mu=(P'-P)^\mu$, $U(P,S)$ is the spinor, and $M$ is the system mass.
In the Drell-Yan frame with $q^+ = 0$, the light-front four momenta are defined as :
\be
&&P=(P^+,P_{\perp},P^-)=\bigg(P^+,0,\frac{M^2}{P^+}\bigg) \, ,\nonumber\\
&&P'=(P'^+,P'_{\perp},P'^-)=\bigg(P^+,q_{\perp},\frac{q_{\perp}^2+M^2}{P^+}\bigg) \, ,\nonumber\\
&&q=P'-P=\bigg(0,q_{\perp},\frac{q_{\perp}^2}{ P^+}\bigg) \, ,
\ee
By calculating the $(++)$ component of energy-momentum tensor, one can obtain
\be\label{AB}
\langle P+q, \uparrow|T_i^{++}(0)|P, \uparrow \rangle &=&{2 (P^+)^2} A_i(Q^2), \\
\langle P+q, \uparrow|T_i^{++}(0)|P, \downarrow \rangle &=& -{2 (P^+)^2}\frac{(q^1_\perp - i q^2_\perp)}{2M}B_i(Q^2)
\label{gff1}
\ee
with $Q^2=-q^2=\vec{q}_\perp^2$. $A_i(Q^2)$ and $B_i(Q^2)$ are very similar to the Dirac and Pauli form factors which are obtained from the helicity non-flip and helicity flip matrix elements of the vector current. Meanwhile, the form factors $C_i(Q^2)$ and $\bar{C}_i(Q^2)$ can be extracted from the helicity flip matrix elements of $T^{-\perp}$ and $T^{+-}$ components. The matrix elements of $T^{-\perp}$ and $T^{+-}$ give:
\be\label{C}
&&\langle P+q, \uparrow| T_i^{-2}(0)| P, \downarrow\rangle +
\langle P+q, \downarrow| T_i^{-2}(0)| P, \uparrow\rangle \nonumber\\
&&= \frac{1}{P^+} \big[2 A_i(Q^2) M^2 - \left(B_i(Q^2) -4 C_i(Q^2)\right) Q^2  \big] \frac{-i  (q^2_{\perp})^2  }{2M},
\label{eqm2lhs}\ee
\eq\label{Cbar}
&&\langle P+q, \uparrow| T_i^{+-}(0)| P, \downarrow\rangle +
\langle P+q, \downarrow| T_i^{+-}(0)| P, \uparrow\rangle \nonumber\\
&&=\big[A_i(Q^2)(2M)-B_i(Q^2)\frac{Q^2}{M}+C_i(Q^2)\frac{4Q^2}{M}+\bar{C}_i(Q^2)(4M)\big](-iq^2_{\perp}) \,.
\en
%
We consider the energy momentum tensor of a free quark inside the proton
to evaluate the form factors:
\be
T^{\mu\nu}=\frac{i}{2}[\bar{\psi}\gamma^{\mu}(\overrightarrow{\partial}^{\nu}\psi)-\bar{\psi}\gamma^{\mu}\overleftarrow{\partial}^{\nu}\psi],
\ee
where $\psi$ is the quark field. Using the two particle Fock states, Eq.~(\ref{state}), and the LFWFs given in Eq.~(\ref{WF}), we evaluate the matrix elements of $T^{++}$, $T^{-\perp}$, and $T^{+-}$ and extract the GFFs $A(Q^2)$, $B(Q^2)$, $C(Q^2)$, and $\bar{C}(Q^2)$ from Eqs.~(\ref{AB})-(\ref{Cbar}). We obtain,
%
\begin{align}
A^q(Q^2) &= \mathcal{I}_1^q(Q^2),\label{a}\\
B^q(Q^2) &= \mathcal{I}_2^q(Q^2),\label{b}\\
C^q(Q^2) &= -\frac{1}{4Q^2}\big[2M^2\mathcal{I}_1^q(Q^2)-Q^2\mathcal{I}_2^q(Q^2)-\mathcal{I}_3^q(Q^2)\big],\label{c}\\
\bar{C}^q(Q^2)
&=-\frac{1}{4M^2}\big[\mathcal{I}_3^q(Q^2)-\mathcal{I}_4^q(Q^2)\big],\label{cbar}
\end{align}
where the explicit expressions of the structure integrals $\mathcal{I}_i^q(Q^2)$ are given by 
\begin{align}
\mathcal{I}_1^q(Q^2)=&~
\int dx x \Big[{N_1}^2 x^{2a_1}(1-x)^{2b_1+1}+{N_2}^2x^{2a_2-2}(1-x)^{2b_2+3}\frac{1}{M^2}
\left(\frac{\kappa^2}{\log(1/x)}-\frac{Q^2}{4}\right)\Big]\exp\Big[-\frac{\log(1/x)}{\kappa^2}\frac{Q^2}{4}\Big],\label{I1}\\
\mathcal{I}_2^q(Q^2)=&~
2\int dx~ {N_1}{N_2} x^{a_1+a_2}(1-x)^{b_1+b_2+2}\exp\Big[-\frac{\log(1/x)}{\kappa^2}\frac{Q^2}{4}\Big],\label{I2}\\
\mathcal{I}_3^q(Q^2)=&~
2\int dx {N_2}{N_1} x^{a_1+a_2-2} (1-x)^{b_1+b_2+2} \nn\\
&\times \Big[
\frac{4(1-x)^2 \kappa^2}{\mathrm{log}(1/x)} + Q^2(1-x)^2-4m^2 
\Big]\exp\Big[-\frac{\log(1/x)}{\kappa^2}\frac{Q^2}{4}\Big],\label{I3}\\
\mathcal{I}_4^q(Q^2)=&
-2 \int dx~ {N_2}{N_1} x^{a_1+a_2-2}(1-x)^{b_1+b_2+2}\Big[\frac{\kappa^2(1-x)^2}{\log(1/x)}+\frac{Q^2(1-x)^2}{4}+m^2\Big]\exp\Big[-\frac{\log(1/x)}{\kappa^2}\frac{Q^2}{4}\Big].\label{I4}
\end{align}
The expressions of $\mathcal{I}_i^q(Q^2)$ in terms of the overlap of
the LFWFs are listed in the Appendix. Although, the GFFs $A(Q^2)$ and $B(Q^2)$ are well defined for all region of $Q^2$ in our model, it can be noticed from Eq.~(\ref{c}) that the form factor $C(Q^2)$ is not well defined at $Q^2=0$. Thus, it is not reachable for the region of small $Q^2$. We observe that the partial cancellation between $2M^2\mathcal{I}_1^q(Q^2)$ and $\mathcal{I}_3^q(Q^2)$ in Eq.~(\ref{c}) provides a nonzero constant (coefficient of $(Q^2)^0$), which brings the $1/Q^2$ singularity in the form factor $C(Q^2)$. Since, the GFFs are physical observables and therefore finite, we emphasize that the singularity appearing in Eq.~(\ref{c}) is a shortcoming of the actual holographic model used in this work but not an artifact of the light-front framework. However, following the approach adopted in Ref.~\cite{Anikin:2019kwi}, we approximately fit this form factor in the large $Q^2~(\ge 0.1~{\rm GeV^2})$ region and, then, employ an analytical continuation of our result to $0\le Q^2<0.1$ GeV$^2$ domain. 
\begin{figure}[!htp]
\centering
(a)\includegraphics[width=8.0cm]{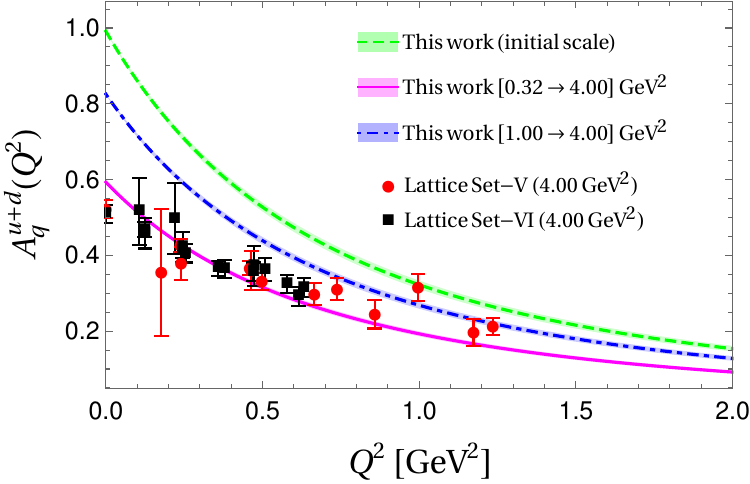}
(b)\includegraphics[width=8.0cm]{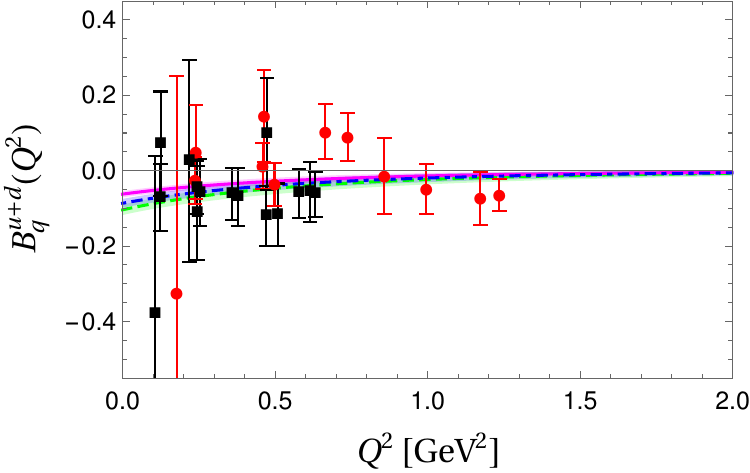}
\caption{The plots of GFFs (a) $A^{u+d}(Q^2)$, and (b) $B^{u+d}(Q^2)$ as functions of $Q^2$. The dashed green lines with green bands are the results at the initial scale, while the solid magenta lines with magenta bands and the dash-dotted blue lines with purple bands represent the results at $\mu^2=4$ GeV$^2$ evolved from the initial scales $\mu_0^2=0.32$ GeV$^2$ and  $\mu_0^2=1.00$ GeV$^2$,  respectively.  The error bands correspond to $2\%$ uncertainty in the model parameters. Our results  are compared with the lattice results (red circle and black square) at scale $\mu^2=4$ GeV$^2$~\cite{Hagler:2007xi}.}
\label{fig:AB}
\end{figure}
\begin{figure}[!htp]
\centering
(a)\includegraphics[width=8.0cm]{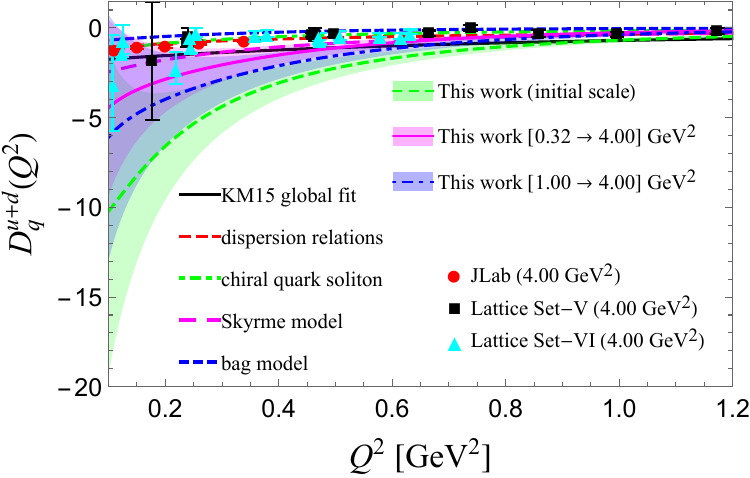}
(b)\includegraphics[width=8.0cm]{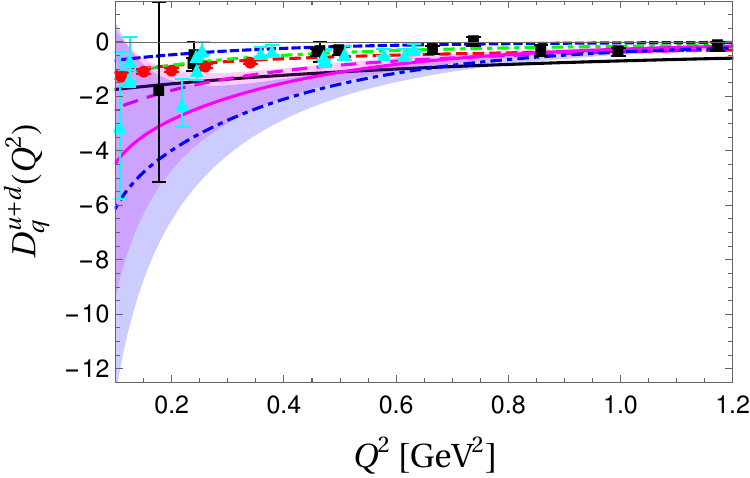}
\caption{The plots of GFF $D^{u+d}_q(Q^2)=4C^{u+d}_q(Q^2)$ as a function of $Q^2$. The dashed green line with green band, the solid magenta line with magenta band and the dash-dotted blue line with purple band in plot (a) represent the results at the initial scale and at scale $\mu^2=4$ GeV$^2$ evolved from the initial scales $\mu_0^2=0.32$ GeV$^2$ and  $\mu_0^2=1.00$ GeV$^2$, respectively. We show the result at $\mu^2=4$ GeV$^2$ evolved from $\mu_0^2=0.32$ GeV$^2$ and  $\mu_0^2=1.00$ GeV$^2$ separately in plot (b). The error bands correspond to $2\%$ uncertainty in the model parameters. The red circles are the experimental data from the Jefferson Lab~\cite{Burkert:2018bqq} and the cyan triangles and black squares correspond to the lattice results~\cite{Hagler:2007xi}. Our results  are compared with KM15 global fit~\cite{Kumericki:2015lhb} (solid black), dispersion relations~\cite{Pasquini:2014vua} (dashed red), chiral quark soliton~\cite{Goeke:2007fp} (dash-dotted green), Skyrme model~\cite{Cebulla:2007ei} (big dashed magenta), and bag model~\cite{Ji:1997gm} (dashed blue).
}
\label{fig:C}
\end{figure}
\begin{figure}[!htp]
\centering
\includegraphics[width=8.0cm]{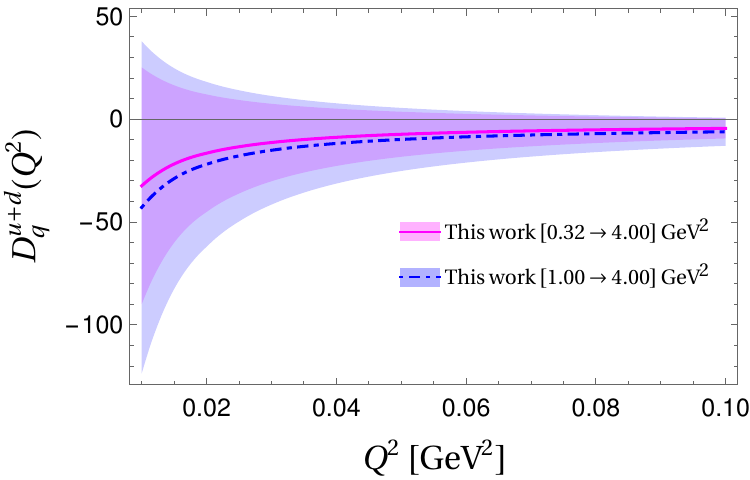}
\caption{The plot of GFF $D^{u+d}_q(Q^2)=4C^{u+d}_q(Q^2)$ at low $Q^2$ region. The solid magenta lines with magenta bands and the dash-dotted blue lines with purple bands represent the results at $\mu^2=4$ GeV$^2$ evolved from the initial scales $\mu_0^2=0.32$ GeV$^2$ and  $\mu_0^2=1.00$ GeV$^2$,  respectively. The error bands correspond to $2\%$ uncertainty in the model parameters.}
\label{fig:C2}
\end{figure}
\begin{figure}[!htp]
\centering
(a)\includegraphics[width=8.0cm]{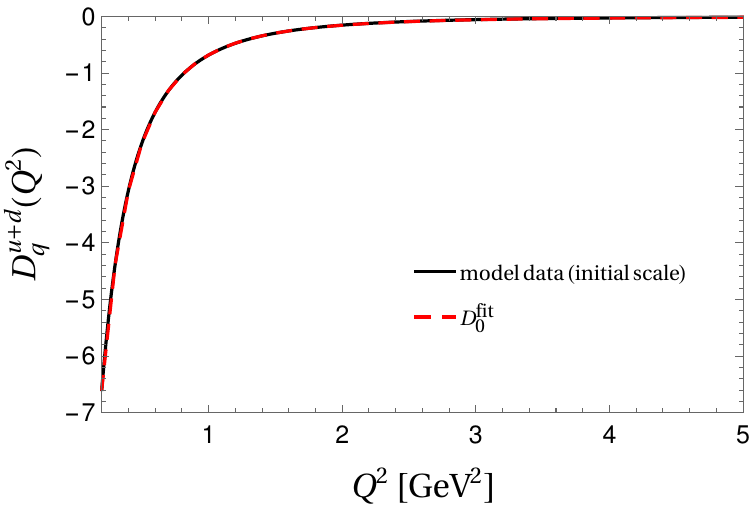}
(b)\includegraphics[width=8.0cm]{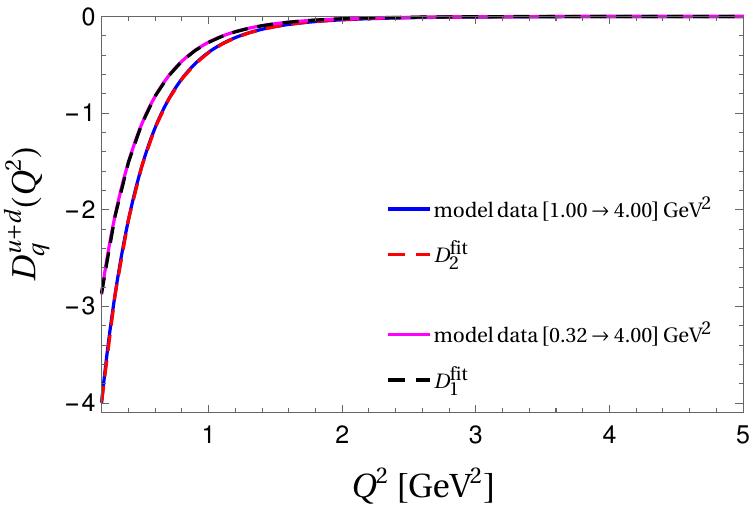}
\caption{Comparison between the model data (black lines) and the fitting function in Eq.~(\ref{fit}) for $D^{u+d}_q(Q^2)$ (dashed red). Left panel is before QCD evolution at the initial scale and right panel is after the QCD evolution at $\mu^2=4$ GeV$^2$ evolved from $\mu_0^2=0.32$ GeV$^2$ and  $\mu_0^2=1.00$ GeV$^2$, respectively. The solid lines represent the actual model results, while the dashed lines correspond to the fits using Eq.~(\ref{fit}) with the parameters given in Table~\ref{table1}. }
\label{fig:Cfit}
\end{figure}

Similar to
the electromagnetic densities, one can interpret the two-dimensional Fourier transform of the GFF $A(Q^2)$ as the longitudinal momentum density in the transverse plane~\cite{Abidin:2008sb,Mondal:2015fok,Chakrabarti:2015lba,Selyugin:2009ic,Kumar:2017dbf,Kumano:2017lhr,Mondal:2016xsm,Kaur:2018ewq}, while the GFF $B(Q^2)$ provides an anomalous contribution to the longitudinal momentum
densities in a transversely polarized target.
In Fig.~\ref{fig:AB}, we show the GFFs $A^{u+d}(Q^2)$ and $B^{u+d}(Q^2)$. 
Our results are compared with the lattice QCD prediction~\cite{Hagler:2007xi}.
lattice data are available at the scale $\mu^2=4$ GeV$^2$. Thus, in order to compare with lattice QCD prediction, we obtain $A^q(Q^2)$ and $B^q(Q^2)$ at the relevant scale by performing the QCD evolution of the integrands of Eqs.~(\ref{I1}) and (\ref{I2}), which represent GPDs $xH^q(x,Q^2)$ and $xE^q(x,Q^2)$, respectively. We adopt the Dokshitzer-Gribov-Lipatov-Altarelli-Parisi (DGLAP) equations \cite{Dokshitzer:1977sg,Gribov:1972ri,Altarelli:1977zs} 
of QCD with next-to-next-to-leading order (NNLO) for the scale evolution. Explicitly, we evolve the GPDs from the model's scale $\mu^2_0=0.32$ GeV$^2$  to the relevant lattice scale $\mu^2=4$ GeV$^2$ using the higher order perturbative parton evolution toolkit (HOPPET)~\cite{Salam:2008qg}. We find that after QCD evolution, $A^{u+d}(Q^2)$ and $B^{u+d}(Q^2)$ are  consistent with the lattice QCD results. We also choose a higher initial scale $\mu_0^2=1.0$ GeV$^2$ and perform the analysis of our results. We observe that the lower initial scale is preferable after comparing our results with the lattice QCD predictions.

\begin{table}[ht]
\caption{Parameters for the fitted function $D^{q}_{\mathrm{fit}}$ given in Eq.~(\ref{fit}) at high $Q^2$. $D^{\mathrm{fit}}_0$ is the distribution at the initial scale, while $D^{\mathrm{fit}}_1$ and $D^{\mathrm{fit}}_2$ are the distributions at the scale $\mu^2=4$ GeV$^2$ evolved from the initial scale $\mu_0^2=0.32$ GeV$^2$ and  $\mu_0^2=1.00$ GeV$^2$, respectively. Here, $[\mu_0^2 \to \mu^2]$ represents the evolution from $\mu_0^2$ to $\mu^2$. }
\centering
\begin{tabular}[t]{lccccc}
\toprule\hline
Parameters~~ & ~~$\mu^2(\mathrm{GeV}^2)$~~ &~~$a_q$~~ & ~~$b_q$~~ & ~~$c_q$~~ &\\
\midrule\hline
~~~$D_{0}^{\mathrm{fit}} $ &~initial scale~ &~$-18.8359$~ & ~$-2.2823$~ &  ~$2.7951$~&\\
~~~$D_{1}^{\mathrm{fit}} $ &~$[0.32\to 4]$~ &~$-5.5861$~ &~$-0.29724$~ & ~$11.6641$~ & \\
~~~$D_{2}^{\mathrm{fit}} $ &~$[1.00\to 4]$~ &~$-7.77884$~ &~$0.291081$~ & ~$11.884$~ & \\
\hline\hline
\end{tabular}
\label{table1}
\end{table}%

The form factor $4C^{u+d}(Q^2)$, also known as $D$-term is displayed in Fig.~\ref{fig:C}, where after scale evolution, we find that the qualitative behavior of our result is compatible with lattice~\cite{Hagler:2007xi} and the experimental data from JLab~\cite{Burkert:2018bqq} as well as other theoretical predictions from the KM15 global fit~\cite{Kumericki:2015lhb}, dispersion relation~\cite{Pasquini:2014vua}, $\chi$QSM~\cite{Goeke:2007fp}, Skyrme model~\cite{Cebulla:2007ei}, and bag model~\cite{Ji:1997gm}. The error bands in our results are due to a $2\%$ uncertainty in the model parameters. The uncertainty in the $D_q(Q^2)$ reflects that the form factor is highly sensitive to our model parameters. The actual uncertainty of the model parametrization in the low $Q^2$ region, $0.01\le Q^2\le 0.1$ GeV$^2$, is shown in Fig.~\ref{fig:C2}.

\begin{figure}[!htp]
\centering
\includegraphics[width=8.0cm]{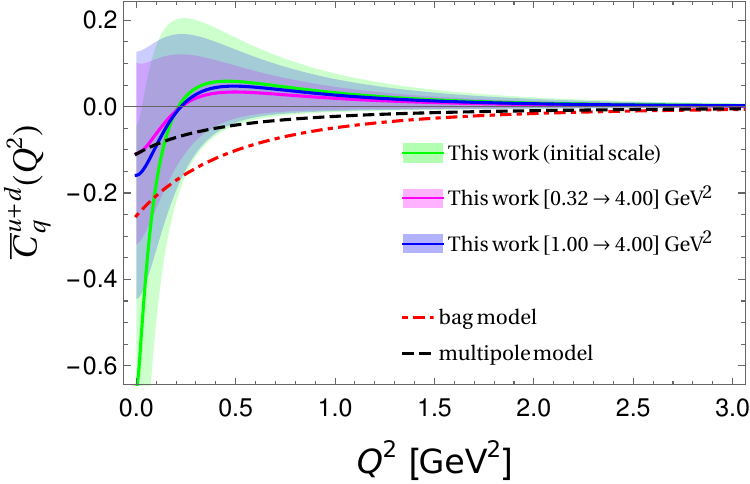}
\caption{The GFF  $\bar{C}^{u+d}_q(Q^2)$ as a function of $Q^2$. The dashed green lines with green bands are the results at the initial scale, while the solid magenta lines with magenta bands and the dash-dotted blue lines with purple bands represent the results at $\mu^2=4$ GeV$^2$ evolved from the initial scale $\mu_0^2=0.32$ GeV$^2$ and  $\mu_0^2=1.00$ GeV$^2$, respectively. The error bands correspond to $2\%$ uncertainty in the model parameters. Our results  are compared with the bag model~\cite{Ji:1997gm} (dash-dotted red) and the multipole model~\cite{Lorce:2018egm} (dashed black).}
\label{fig:Cbar}
\end{figure}

It turns out that the form factor $D^{u+d}(Q^2)$ can be sufficiently described
by the following multipole function defined as
\be\label{fit}
D^q_{\rm fit}(Q^2)=4C^q_{\mathrm{fit}}(Q^2) = \frac{a_q}{(1+b_qQ^2)^{c_q}},
\ee
where the parameters $a_q$, $b_q$, and $c_q$ are given in the Table~\ref{table1}. In Fig.~\ref{fig:Cfit}, we compare the model data of the $D$-term form factor and the multipole function given in Eq.~(\ref{fit}).
The form factor $\bar{C}(Q^2)$ in the quark-diquark model is illustrated in Fig.~\ref{fig:Cbar}. In accord with the bag model~\cite{Ji:1997gm} and the multipole model~\cite{Lorce:2018egm}, $\bar{C}(Q^2)$ in the present model is negative at low $Q^2$~($<0.22$ GeV$^2$), however, we observe a distinctly different behavior in the region of $Q^2>0.22$ GeV$^2$, where it exhibits positive distribution, while in other models $\bar{C}(Q^2)$ is always negative. The positive distribution decreases with QCD evolution.
 We remark that the $D$-term and $\bar{C}(Q^2)$ form factors presented in Figs.~\ref{fig:C}, \ref{fig:C2} and \ref{fig:Cbar}, respectively, directly follow from Eqs.~(\ref{c}) and (\ref{cbar}) computed using the model wave functions and are not the results of any independent parameterization.

We present a comparison of the GFFs at $Q^2=0$ with those of the various phenomenological models, lattice QCD and existing experimental data for $D_q(0)$ in Table~\ref{compare}. For the $A_q(0)$ and $J_q(0)$ form factors, our estimation is in more or less agreement with the predictions of Refs.~\cite{Hagler:2003jd,Gockeler:2003jfa,Bratt:2010jn,Hagler:2007xi,Brommel:2007sb,Deka:2013zha,Dorati:2007bk,Lorce:2018egm} at re-normalization scale of $\mu^2=4$ GeV$^2$. The QCD sum rule (QCDSR) (I $\&$ II) gives a  higher value of $A_q(0)$, since the scale is relatively low $\mu^2=1$ GeV$^2$~\cite{Azizi:2019ytx}.  Note that in the $\chi$QSM  and Skyrme models, there are only quarks and antiquarks to carry the nucleon's angular momentum and they must carry 100$\%$  of it and thus 2\,$J_q (0) = A_q(0)$ = 1. The results in AdS/QCD models are presented at model scale, where $u$ and $d$ quarks together carry $\sim 90\%$ of nucleon momentum. For the form factor $D_q(0)$, our extrapolated value is overestimated when we compare it with lattice QCD results~\cite{Gockeler:2003jfa,Bratt:2010jn,Hagler:2007xi} and the predictions of Refs.~\cite{Azizi:2019ytx,Goeke:2007fp,Anikin:2017fwu} as well as the experimental data from JLab~\cite{Burkert:2018bqq} but they differ from the other predictions presented in the
Table~\ref{compare}. Our predictions for  $\bar C_q (0)$ accords with IFF \cite{Lorce:2018egm} and the asymptotical value at three loops level reported in Ref.~\cite{Hatta:2018sqd}, however, substantially differ from the predictions of QCDSR~\cite{Azizi:2019ytx} and IP~\cite{Polyakov:2018exb}. Note that the intrinsic spin sum rule for a transversely polarized nucleon not only involves  the form factors $A_q$ and $B_q$ but also $\bar C_q $~\cite{Leader1,HKMR,hari,Hatta:2018sqd}.
%
\begin{table}[t]
\caption{The GFFs of valence quark combination at $Q^2=0$ compared with other predictions and JLab data. The Skyrme 
and $\chi$QSM models predictions have been obtained by considering both the quark and the gluon contributions and these are scale independent. Here, $(\mu_0  \to \mu)$ represents the evolution from the initial scale $\mu_0$ to the final scale $\mu$.}
	\addtolength{\tabcolsep}{2pt}
\begin{tabular}{lccccccccc}
				\toprule\hline
				Approaches/Models ~&~ $A^{u+d}_q(0)$ ~&~ $J_q(0)=\frac{1}{2}[A^{u+d}_q(0)+B^{u+d}_q(0)]$ ~&~ $D^{u+d}_{\rm fit}(0)=4C^{u+d}_{\rm fit}(0)$~&~ $\bar C^{u+d}_q(0)$\\
				\hline\hline
				This work ($\sqrt{0.32}$ GeV $\to$ $2$ GeV)    &  0.593 & 0.269 &  -5.586 & -0.109\\
				This work ($1.00$ GeV $\to$ $2$ GeV)    &  0.825 & 0.369 &  -7.778 & -0.159\\
				LQCD ($2$ GeV)~\cite{Hagler:2003jd}    & 0.675 & 0.34 & -&- \\
				LQCD ($2$ GeV)~\cite{Gockeler:2003jfa} & 0.547 & 0.33 & -0.80 &-\\
				LQCD ($2$ GeV)~\cite{Bratt:2010jn}     & 0.553 & 0.238 & -1.02 &-\\
				LQCD ($2$ GeV)~\cite{Hagler:2007xi}    & 0.520 & 0.213 & -1.07&-\\
				LQCD ($2$ GeV)~\cite{Brommel:2007sb}   & 0.572 & 0.226 & - &- \\
				LQCD ($2$ GeV)~\cite{Deka:2013zha}     & 0.565 & 0.314 & - &-\\  
				$\chi$PT ($2$ GeV)~\cite{Dorati:2007bk}       & 0.538 & 0.24 & -1.44 &- \\
				IFF ($2$ GeV)~\cite{Lorce:2018egm}            & 0.55 & 0.24 &-1.28 &-0.11\\
				Asymptotic ($\infty$ GeV)~\cite{Hatta:2018sqd}            & - & 0.18 &- &-0.15\\
				QCDSR-I (1 GeV)~\cite{Azizi:2019ytx}       & 0.79 & 0.36 & -1.832 &-2.1 $\times 10^{-2}$ \\
				QCDSR-II (1 GeV)~\cite{Azizi:2019ytx}      & 0.74 & 0.30 & -1.64 &-2.5 $\times 10^{-2}$\\
				Skyrme \cite{Cebulla:2007ei}        & 1 & 0.5 & -3.584 &-\\
				Skyrme \cite{Kim:2012ts}            & 1 & 0.5 & -2.832 &-\\				
				$\chi$QSM \cite{Goeke:2007fp}       & 1 & 0.5 & -1.88 &-\\
				$\chi$QSM \cite{Jung:2014jja}       & 1 & 0.5 & -4.024 & \\
				$\chi$QSM \cite{Wakamatsu:2007uc}   & - & -& -3.88 &- \\
								AdS/QCD Model I \cite{Mondal:2015fok}       & 0.917 & 0.415& - &- \\
								AdS/QCD Model II \cite{Mondal:2015fok}       & 0.8742 & 0.392 & - &- \\
				LCSR-LO \cite{Anikin:2019kwi}       & -&- & -2.104 &- \\
				KM15 fit \cite{Anikin:2017fwu}      &-&-& -1.744  &-\\
				DR \cite{Pasquini:2014vua}          & - & - & -1.36 &-\\
				JLab data \cite{Burkert:2018bqq}    &- &-& $ -1.688 $ &-\\
				IP \cite{Polyakov:2018exb}          & -&-&-&$ 1.4 \times 10^{-2}$\\
			
      \hline\hline		
	\end{tabular}
	\label{compare}
\end{table} 
\begin{table}[ht]
\caption{The  mechanical properties: pressure, energy density, and mechanical radius of nucleon.}
\begin{tabular}[t]{ l c c c }
 \toprule\hline
Approaches/Models ~&~ $p_0$  [GeV/fm$^3$]                                     ~&~ ${\cal E}$ [GeV/fm$^3$]                                  ~&~ $\langle {r^2_{\text{mech}}\rangle}$ [fm$^{2}$]\\
\hline\hline
  This work ($\sqrt{0.32}$ GeV $\to$ $2$ GeV) &$0.29$ & $ 3.21$ & $0.74$\\
       This work ($1.00$ GeV $\to$ $2$ GeV) &$0.40$ & $ 4.58$ & $0.74$\\
      QCDSR set-I (1 GeV)~\cite{Azizi:2019ytx} &$0.67$ & $1.76$ & $0.54$\\ 
      QCDSR set-II (1 GeV)~\cite{Azizi:2019ytx} & $0.62$ & $1.74$ & $0.52$\\
   Skyrme model~\cite{Cebulla:2007ei} & 0.47 & 2.28 & -\\ 
   modified Skyrme model~\cite{Kim:2012ts} & 0.26 & 1.445 & -\\
     $\chi$QSM~\cite{Goeke:2007fp} & 0.23 & 1.70 & - \\
     Soliton model~\cite{Jung:2014jja} & 0.58 & 3.56 & -\\
         LCSM-LO~\cite{Anikin:2019kwi} & 0.84 & 0.92 & 0.54\\
\hline\hline
\end{tabular}

	\label{table2}
\end{table}
\section{Mechanical properties}\label{properties}
The pressure and the energy density in the
center of nucleon  are directly related to the GFFs as \cite{Polyakov:2018zvc}
\begin{align}\label{pressure}
p_0 &=-\frac{1}{24\pi^2 M_n} \int^{\infty}_{0} {\rm d}Q^2~ Q^3~ {D}(Q^2),\nonumber\\
{\cal E} &=\frac{M_n}{4\pi^2} \int^{\infty}_{0} {\rm d} Q^2
\left( A(Q^2) +
\frac{Q^2}{4M^2_n} D(Q^2) \right),
\end{align}
respectively, while the mechanical radius can be obtained by
\begin{eqnarray}
\label{Rmech}
\langle r^2_{\text{mech}}\rangle=6 D_{\rm fit} (0) \Big[ \int^{\infty}_{0} {\rm d}Q^2~ D(Q^2)\Big]^{-1}.
\end{eqnarray}
Here, $M_n$ denotes the mass of nucleon. We compute the pressure, energy densities in the proton and the mechanical radius using the GFFs evaluated in the quark-diquark model.  Our results  on the mechanical quantities $ p_0 $, $ {\cal E} $, and $\langle r^2_{\text{mech}}\rangle$ of the proton compared to other existing theoretical predictions are shown in the Table~\ref{table2}. It can be seen from Table \ref{table2} that our prediction on the $ p_0 $ is underestimated but comparable with $\chi$QSM model~\cite{Goeke:2007fp} and the modified Skyrme model~\cite{Kim:2012ts}. 
Note that the available theoretical predictions  differ considerably from each other. Our results on $ {\cal E} $ in quark-diquark model is close to the predictions of the soliton model~\cite{Jung:2014jja}, but overestimated compared to the other presented predictions~~\cite{Azizi:2019ytx,Cebulla:2007ei,Kim:2012ts,Goeke:2007fp,Jung:2014jja}. However, different approaches/models demonstrate considerable deviations from each other while predicting $ {\cal E}$. Meanwhile, our predictions on the mechanical radius, $\langle r^2_{\text{mech}}\rangle$, is larger than the
prediction of Refs.~\cite{Azizi:2019ytx,Anikin:2019kwi}.

\begin{figure}[!htp]
\centering
(a)\includegraphics[width=8.0cm]{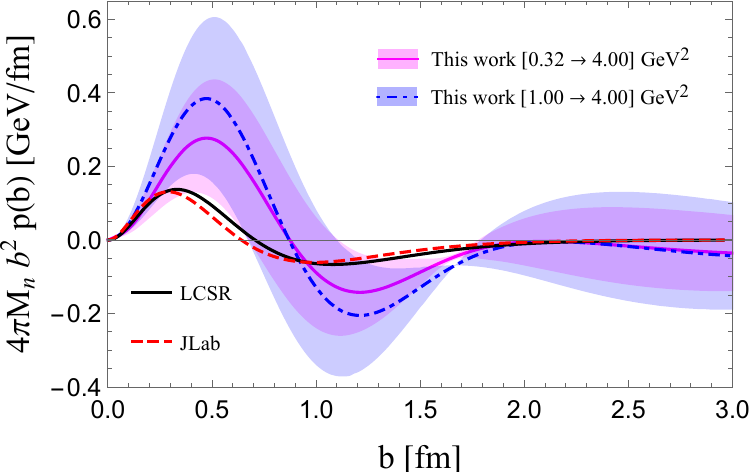}
(b)\includegraphics[width=8.0cm]{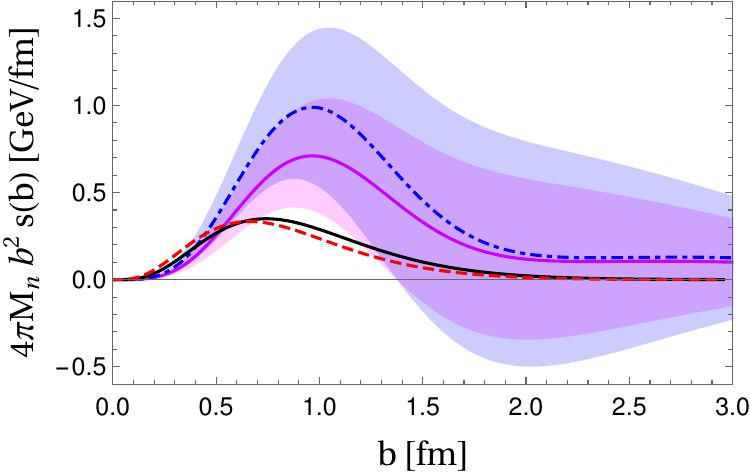}
\caption{Plots of (a) the pressure distribution $4\pi M_nb^2p(b)$, and (b) the shear force distribution $4\pi M_nb^2s(b)$ as a function of $b$. Our results are compared with results based on LCSR evaluated in Ref.~\cite{Anikin:2019kwi} (black line)  and using the fitting function of $D(Q^2)$ based on experimental data~\cite{Burkert:2018bqq} (red dashed line). The solid magenta lines with magenta bands and the dash-dotted blue lines with purple bands represent the results at $\mu^2=4$ GeV$^2$ evolved from the initial scale $\mu_0^2=0.32$ GeV$^2$ and  $\mu_0^2=1.00$ GeV$^2$, respectively.
}
\label{fig:pressure}
\end{figure}
\subsection{Pressure and shear force distributions}
The distributions of pressure and shear forces inside the nucleon are given by
 \begin{align}
\label{Pre-fun}
p(b)=&\frac{1}{6M_n}\frac{1}{b^2}\frac{d}{db} b^2 \frac{d}{db} \tilde{D}(b),\nonumber\\
s(b)=&-\frac{1}{4M_n} b \frac{d}{db} \frac{1}{b} \frac{d}{db} \tilde{D}(b),
\end{align}
where
\be
\tilde{D}(b)
=\int \frac{{\rm d^2}\vec{q}_{\perp}}{(2\pi)^2}~e^{i\vec{q}_{\perp}.\vec{b}_{\perp}} D(Q^2).
\ee
Here, $b=|{\vec{b}_{\perp}}|$ represents the impact parameter. The pressure distribution $p(b)$ must satisfy the stability condition, also known as the von Laue condition~\cite{VonL},
\begin{eqnarray}
\label{Laue}
\int_{0}^{\infty} db\, b^2\, p(b) = 0.
\end{eqnarray}
This is a consequence of the energy momentum tensor conservation and allows us to understand how the internal forces balance inside a composed system
\cite{Polyakov:2018zvc,Polyakov:2002yz}. We illustrate the distribution $b^2p(b)$ as a function of $b$ in Fig.~\ref{fig:pressure}(a). We compare our result with the distribution evaluated in leading order light-cone sum rule~\cite{Anikin:2019kwi} and the distribution obtained from the fitting functions of the experimental data for $D(Q^2)$ at JLab~\cite{Burkert:2018bqq}. The distribution must have at least one
node to comply with the von Laue condition, Eq.~(\ref{Laue}). It can be noticed that the distribution has a positive core and a negative tail. This pattern ensures the mechanical stability arguments: the repulsive
forces are required in the inner domain to prevent collapse and the attractive forces in the outer region to bind the system. Our pressure distribution crosses the zero-line
(zero-crossing) near $0.9$ fm (central line), whereas this zero-crossing appears at $0.7$ fm in the result based on light-cone sum rule~\cite{Anikin:2019kwi} and near $0.6$ fm in the JLab distribution~\cite{Burkert:2018bqq}. Overall, the qualitative behavior of the pressure distribution evaluated in the quark-diquark model is found to be in agreement with the experimental observation~\cite{Burkert:2018bqq} as well as other theoretical  predictions~\cite{Shanahan:2018nnv,Anikin:2019kwi,Goeke:2007fp,Kim:2012ts,Jung:2014jja,Cebulla:2007ei}. The shear force distribution, $b^2s(b)$ has been displayed in Fig.~\ref{fig:pressure}(b). $s(b)$ has a connection to surface tension and surface energy, which are positive in stable hydrostatic systems~\cite{Polyakov:2018zvc}. We find that $s(b)$ (central line) is positive in all region of $b$. The positivity of this distribution was observed in all studies so far. We again notice that the qualitative nature of our result is in accordance with other approaches~\cite{Shanahan:2018nnv,Anikin:2019kwi,Goeke:2007fp,Kim:2012ts,Jung:2014jja,Cebulla:2007ei}.
\begin{figure}[!htp]
\centering
(a)\includegraphics[width=8.0cm]{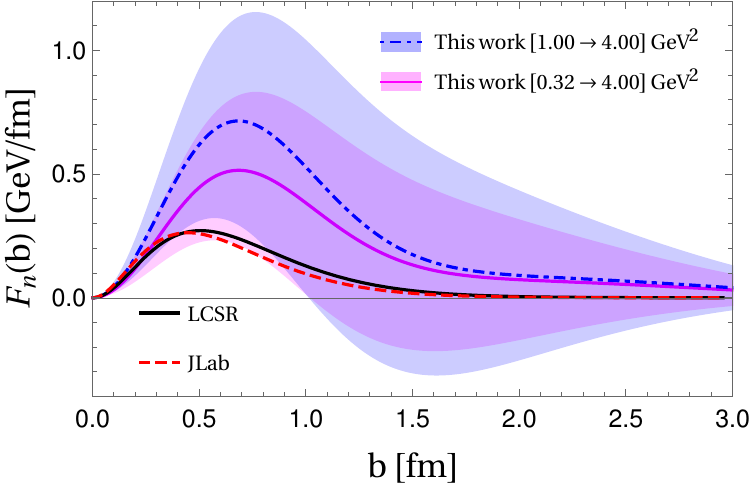}
(b)\includegraphics[width=8.0cm]{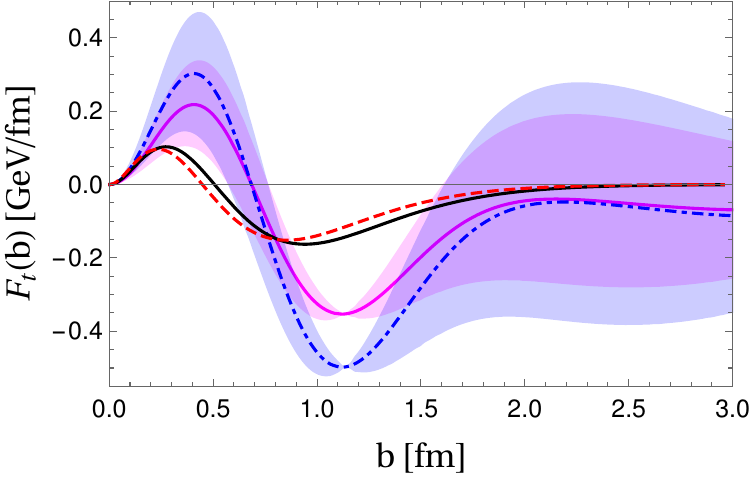}
\caption{Plots of (a) the normal forces  $F_n$, and (b) the tangential forces $F_t$ as a function of $b$. Our results are compared with results based on LCSR evaluated in Ref.~\cite{Anikin:2019kwi} (black line)  and using the fitting function of $D(Q^2)$ based on experimental data~\cite{Burkert:2018bqq} (red dashed line). The solid magenta lines with magenta bands and the dash-dotted blue lines with purple bands represent the results at $\mu^2=4$ GeV$^2$ evolved from the initial scale $\mu_0^2=0.32$ GeV$^2$ and  $\mu_0^2=1.00$ GeV$^2$, respectively.}
\label{fig:force}
\end{figure}

On the other hand, the spherical shell of radius $b$ in the nucleon experiences the normal and tangential forces: $F_n$ and $F_t$, respectively, which are defined as~\cite{Anikin:2019kwi}
\begin{align}
\label{Fn-Ft}
F_n(b)=&4\pi M_n b^2 \left( p(b) + \frac{2}{3} s(b)\right),
\nonumber\\
F_t(b)=&4\pi M_n b^2 \left( p(b) - \frac{1}{3} s(b)\right).
\end{align}
We show the estimated normal and tangential
forces for the valence quark combination in Fig~\ref{fig:force}(a) and Fig~\ref{fig:force}(b), respectively. One notices that $F_n(b)$ is always positive, whereas $F_t(b)$  has
a positive core (repulsive force) surrounded by a negative tail (attractive force)  with a zero-crossing near $b\sim 0.7$ fm.
The repulsive force has the peak near $b\sim 0.4$ fm, and the maximum of the negative force that is responsible for the binding occurs near $b\sim 1.1$ fm. However, the binding force is stronger than the repulsive force. The qualitative behavior of the forces in the quark-diquark model is fairly consistent with the light-cone sum rule~\cite{Anikin:2019kwi}, estimated distribution from JLab fitting function for $D(Q^2)$~\cite{Burkert:2018bqq}, and chiral quark-soliton~\cite{Goeke:2007fp} model as well.

We also compute the two-dimensional Galilean energy density, radial pressure, tangential pressure, isotropic pressure, and pressure anisotropy, which in Drell-Yan-West frame are defined as~\cite{Lorce:2018egm},
\begin{align}
\mu_a(b) =& M_n \Big{\{} \frac{A_a(b)}{2} + \bar{C}_a(b) + \frac{1}{4M_n^2}\frac{1}{b}\frac{d}{db} \left( b\frac{d}{db}\left[ \frac{B_a(b)}{2} - 4C_a(b) \right)\right] \Big{\}}\nn \\
\sigma_{r,a}(b) =& M_n  \Big{\{}
-\bar{C}_a(b) + \frac{1}{M_n^2} \frac{1}{b} \frac{dC_a(b)}{db} 
\Big{\}} \nn \\
\sigma_{t,a}(b) =&  M_n  \Big{\{}
-\bar{C}_a(b) + \frac{1}{M_n^2} \frac{d^2C_a(b)}{db^2} 
\Big{\}} \label{mp} \\ 
\sigma_{a}(b) =& M_n  \Big{\{}
-\bar{C}_a(b) + \frac{1}{2}\frac{1}{M_n^2} \frac{1}{b} \frac{d}{db}\left(b \frac{dC_a(b)}{db} \right)
\Big{\}} \nn \\
\Pi_a(b)=& M_n  \Big{\{} 
-\frac{1}{M_n^2} b \frac{d}{db}\left(b \frac{dC_a(b)}{db} \right)\Big{\}}\nn,
\end{align}
respectively, where the form factors in position space are given by their Fourier transform:
\be
\chi(b)
=\int \frac{{\rm d^2}\vec{q}_{\perp}}{(2\pi)^2}~e^{i\vec{q}_{\perp}.\vec{b}_{\perp}}\chi(q^2).
\ee

\begin{figure}[!htp]
\centering
(a)\includegraphics[width=8.0cm]{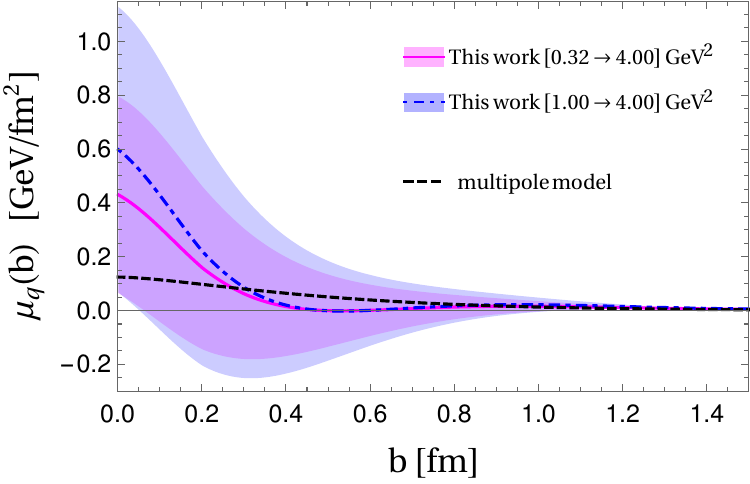}
(b)\includegraphics[width=8.0cm]{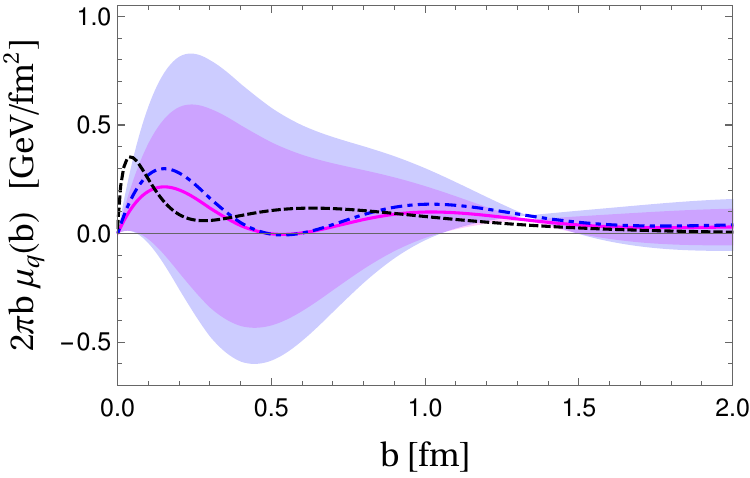}
\caption{Plots of the two-dimensional Galilean energy density (a) $\mu_q(b)$, and (b) $2\pi b \mu_q(b)$. The solid magenta lines with magenta bands and the dash-dotted blue lines with purple bands represent the results at $\mu^2=4$ GeV$^2$ evolved from the initial scale $\mu_0^2=0.32$ GeV$^2$ and  $\mu_0^2=1.00$ GeV$^2$, respectively. Our results are compared with the results in a multipole model (black dashed line)~\cite{Lorce:2018egm}}.
\label{fig:mu}
\end{figure}

\begin{figure}[!htp]
\centering
(a)\includegraphics[width=8.0cm]{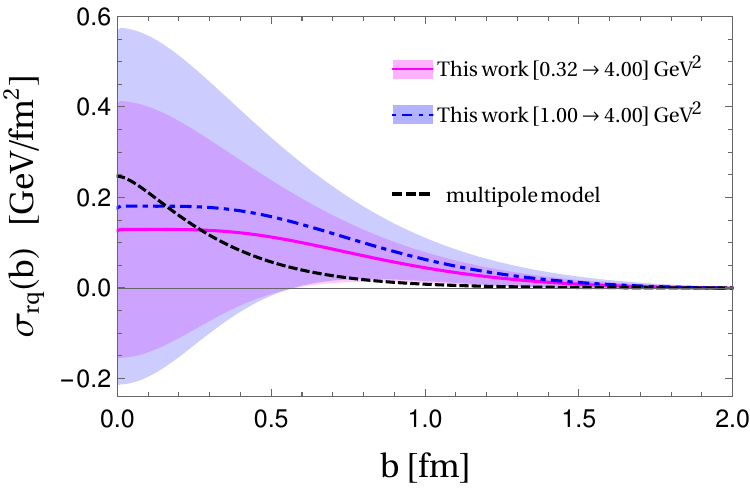}
(b)\includegraphics[width=8.0cm]{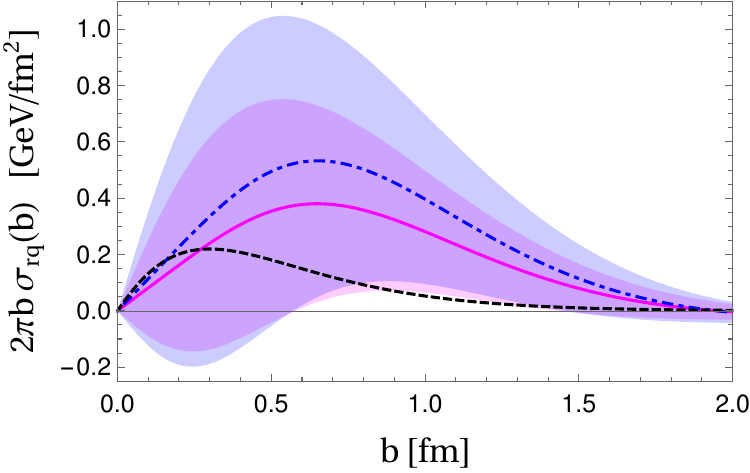}
\caption{Plots of the two-dimensional radial pressure (a) $\sigma_{r,q}(b)$, and (b) $2\pi b \sigma_{r,q}(b)$. The legends are same as mentioned in Fig.~\ref{fig:mu}.}
\label{fig:sigmar}
\end{figure}

\begin{figure}[!htp]
\centering
(a)\includegraphics[width=8.0cm]{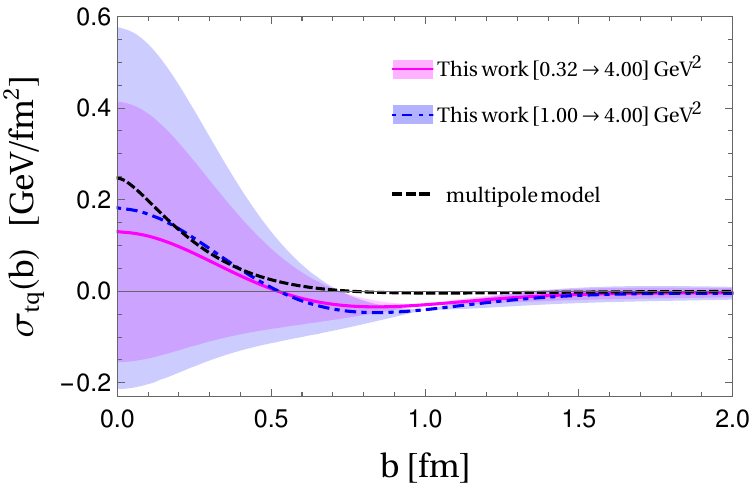}
(b)\includegraphics[width=8.0cm]{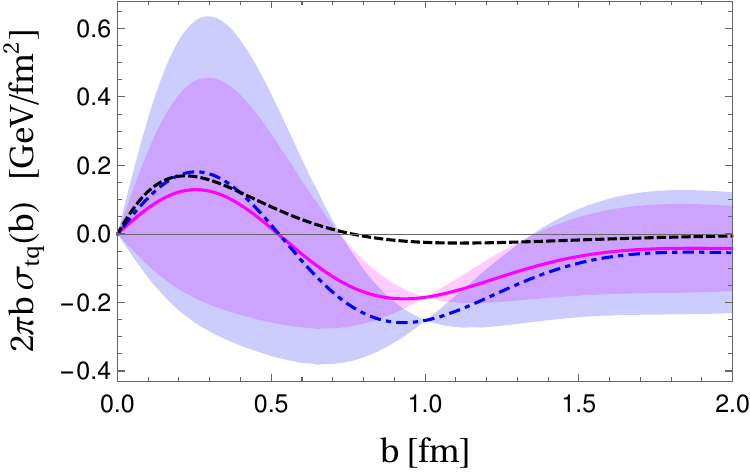}
\caption{Plots of the two-dimensional tangential pressure (a) $\sigma_{t,q}(b)$, and (b) $2\pi b \sigma_{t,q}(b)$. The legends are same as mentioned in Fig.~\ref{fig:mu}.}
\label{fig:sigmat}
\end{figure}

\begin{figure}[!htp]
\centering
(a)\includegraphics[width=8.0cm]{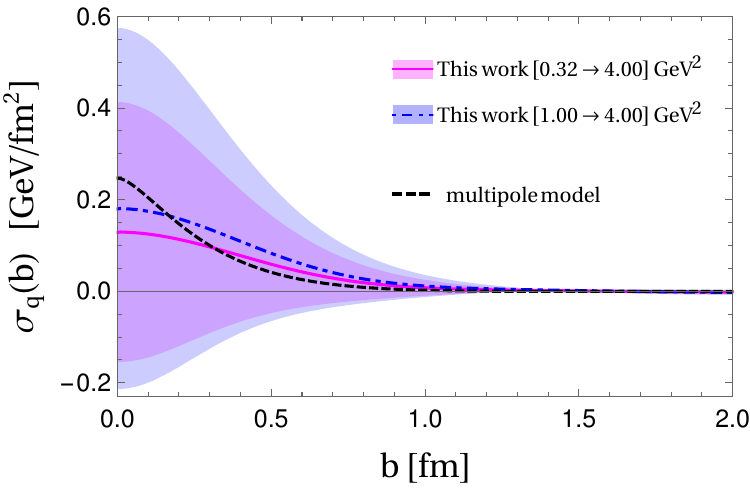}
(b)\includegraphics[width=8.0cm]{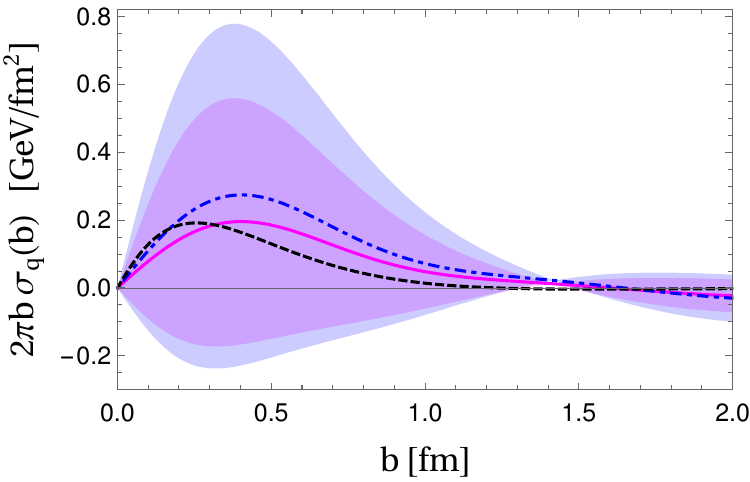}
\caption{Plots of the two-dimensional isotopic pressure (a) $\sigma_{q}(b)$, and (b) $2\pi b \sigma_{q}(b)$. The legends are same as mentioned in Fig.~\ref{fig:mu}.}
\label{fig:sigma}
\end{figure}

\begin{figure}[!htp]
\centering
(a)\includegraphics[width=8.0cm]{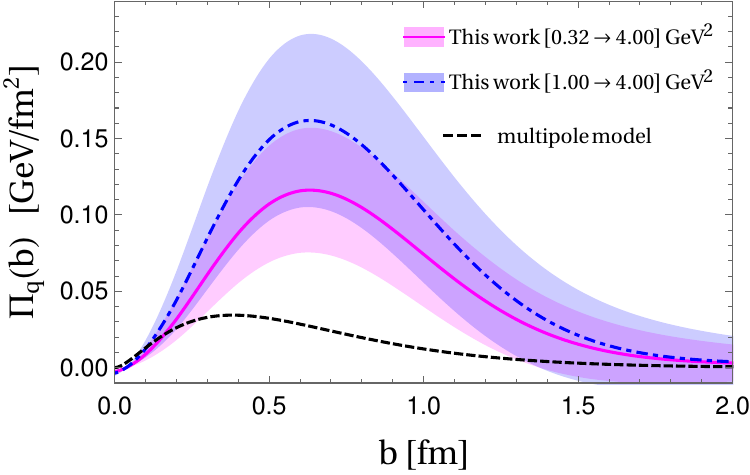}
(b)\includegraphics[width=8.0cm]{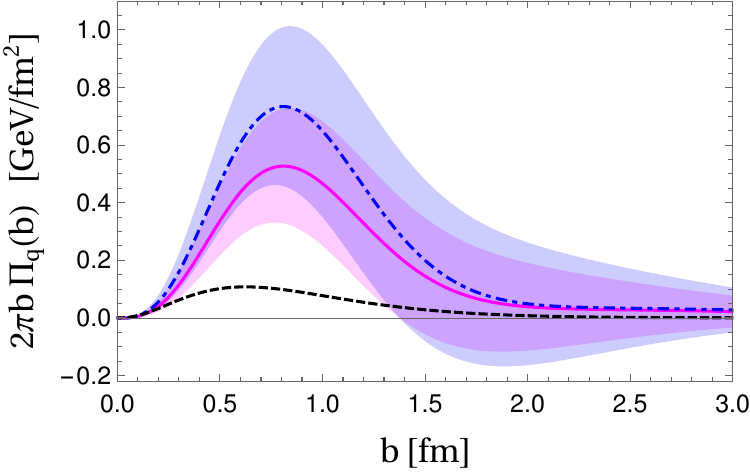}
\caption{Plots of the two-dimensional pressure anisotropy (a) $\Pi_{q}(b)$, and (b) $2\pi b \Pi\sigma_q(b)$. The legends are same as mentioned in Fig.~\ref{fig:mu}.}
\label{fig:pi}
\end{figure}

The distributions defined in Eq.~(\ref{mp}) are illustrated in Figs~\ref{fig:mu}, \ref{fig:sigmar}, \ref{fig:sigmat}, \ref{fig:sigma}, and \ref{fig:pi}. We observe that the energy density $\mu_q(b)$ in Fig.~\ref{fig:mu} and the radial pressure $\sigma_{r,q}(b)$ in Fig.~\ref{fig:sigmar} are always positive having the peaks at center of the proton ($b=0$). On the other hand, in Fig.~\ref{fig:sigmat}, the tangential pressure $\sigma_{t,q}(b)$ is positive at low $b$ with maxima at the center but it shows negative distribution when $b>0.55$ fm. The isotopic pressure $\sigma_q(b)$ in Fig.~\ref{fig:sigma} exhibits a similar behavior as the radial pressure, however, it also shows a slightly negative distribution at large $b$ ($>1.7$ fm).  The pressure anisotropy in Fig.~\ref{fig:pi} vanishes at the center of the proton, as required by spherical symmetry, and is positive anywhere else, indicating that the radial pressure is always larger than the tangential one. Our predictions on
the energy density, radial pressure, tangential pressure, isotropic pressure, and pressure anisotropy  are compared with the results in a simple multipole model~\cite{Lorce:2018egm}. The qualitative behavior of those distributions in Figs~\ref{fig:mu}$-$\ref{fig:pi}, within the error bands, are found be consistent with the multipole model reported in Ref.~\cite{Lorce:2018egm}. We have also observed that the qualitative behavior remains the same going from $\mu_0^2=0.32$ GeV$^2$ to $\mu_0^2=1.0$ GeV$^2$. 

\section{summary}\label{summary}
Gravitational form factors provide us with knowledge on different aspects of the nucleon's structure, e.g., the pressure and energy distributions, the distribution and stabilization of the strong force inside the nucleon as well as quantities related to its geometric shape. One can also get the information on the fractions of the longitudinal momenta carried by the constituents and the total angular momentum from the GFFs. In this paper, we have evaluated the GFFs of the proton in a light-front quark-diquark model with AdS/QCD predicted wave functions. We have showed explicit $Q^2$
behavior of the gravitational form factors in this model and observed that the GFFs $A(Q^2)$ and $B(Q^2)$ are comparable with lattice QCD results~\cite{Hagler:2007xi}. We have found that the GFF $D(Q^2)$ of proton can be described by a multipole fit function. We have observed that the qualitative nature of $D(Q^2)$ in the quark-diquark model is in accord with the experimental data extracted from DVCS process at JLab~\cite{Burkert:2018bqq} and lattice QCD predictions~\cite{Hagler:2007xi}.
We have compared the values of the GFFs at $Q^2=0$ with the existing theoretical predictions and the data from JLab. 
Our results on $A_q (0)$ and $J_q(0)$ were found to be in fair agreement with the lattice QCD and chiral perturbation theory predictions. Meanwhile, our prediction on $D^q(0)$ appeared to be larger while comparing with the JLab data as well as the available theoritical predictions. It should also be noted that there are large discrepancies among the theoretical predictions on the $D_q(0)$. We have extracted the value of $\bar C_q (0)$ which is comparable with that in Refs.~\cite{Lorce:2018egm,Hatta:2018sqd} but larger than those values reported in Refs.~\cite{Polyakov:2018exb,Azizi:2019ytx}. 

We have employed the GFFs to evaluate the pressure and energy density distributions inside the proton as well as the mechanical radius of the
proton and compared them with the existing theoretical predictions. We have observed that the pressure in the center of the proton $p_0$ in quark-diquark model is underestimated with respect to QCDSR~\cite{Azizi:2019ytx}, LCSR-LO~\cite{Anikin:2019kwi}, Soliton model~\cite{Jung:2014jja} and Skyrme model~\cite{Cebulla:2007ei}  but close to that in $\chi$QSM~\cite{Goeke:2007fp} and modified Skyrme model~\cite{Kim:2012ts}, while our prediction on the energy density $\mathcal{E}$ is close to that in the soliton model~\cite{Jung:2014jja}. However, there are large discrepancies among the theoretical predictions on $p_0$ and $\mathcal{E}$. 
Mechanical radius, $\langle r^2_{\text{mech}}\rangle$, has been found to be somewhat larger compared to the existing predictions
provided by LCSR-LO~\cite{Anikin:2019kwi} and QCDSR~\cite{Azizi:2019ytx}.

We have demonstrated the pressure $p(b)$ and shear force $s(b)$ distributions inside the proton in the quark-diquark model. We have noticed that $p(b)$ has  
a positive core and a negative tail, while $s(b)$ is always positive which are consistent with the experimental observation and other theoretical predictions. On the other hand, we also found that the normal force $F_n(b)$ is always repulsive but the tangential force $F_t(b)$ is repulsive near the center but attractive for $b>0.5$ fm. These behaviors are again in fair agreement with experimental observation and other theoretical predictions. We have also computed the two-dimensional Galilean energy density, radial pressure, tangential pressure, isotropic pressure, and pressure anisotropy in this present model, which qualitatively have been found to be consistent with a multipole model.

\section{Acknowledgments}
D. C. and A. M. is supported by
Science and Engineering Research Board under Grant
No. CRG/2019/000895.
C. M. is supported by the National Natural Science Foundation of China (NSFC) under the Grants No.
11850410436 and No. 11950410753. C.M. is also
supported by new faculty start up funding by the Institute of Modern Physics, Chinese Academy of Sciences.  S. N. is supported by the China Postdoctoral Council under the International
Postdoctoral Exchange Fellowship Program. X. Z. is supported by a new faculty start up funding from the Institute of Modern Physics, Chinese Academy of Sciences under the grant No.~Y632030YRC and by Key Research Program of
Frontier Sciences, CAS, Grant No. ZDBS-LY-7020. This work of C. M., S. N., and X. Z. is also supported by the Strategic Priority Research Program of Chinese Academy of Sciences, Grant No. XDB34000000.


\section{APPENDIX: MATRIX ELEMENTS OF $T^{\mu\nu}$}
\subsection{Matrix elements of $T^{++}$}
\begin{align}
&\langle P+q, \uparrow|T_{q}^{++}|P, \uparrow\rangle\nonumber\\
&= 2(P^+)^2\int\frac{d^2k_{\perp}dx}{16\pi^3}  x~ 
\big[\psi_{+\frac{1}{2}}^{\uparrow *}(x,\vec{k'}_{\perp})\psi_{+\frac{1}{2}}^{\uparrow}(x,\vec{k}_{\perp})
+ \psi_{-\frac{1}{2}}^{\uparrow *}(x,\vec{k'}_{\perp})\psi_{-\frac{1}{2}}^{\uparrow}(x,\vec{k}_{\perp})\big] \, \nonumber\\
&= 2(P^+)^2\int dx x \bigg[N_1^2x^{2a_1}(1-x)^{2b_1+1}+N_2^2x^{2a_2-2}(1-x)^{2b_2+3}\frac{1}{M^2}\bigg(\frac{\kappa^2}{\log(1/x)}-\frac{Q^2}{4}\bigg)\bigg]\exp\bigg[-\frac{\log(1/x)}{\kappa^2}\frac{Q^2}{4}\bigg]\nonumber\\
&= 2(P^+)^2~{\mathcal{I}_{1}^q(Q^2)}, 
\end{align}
where $\vec{k'}_{\perp}=\vec{k}_{\perp}+(1-x)\vec{q}_{\perp}$. Using the matrix elements Eq.(\ref{tensor}),
\begin{align}
\langle P+q, \uparrow|T_{q}^{++}|P, \uparrow\rangle=2(P^+)^2A^q(Q^2)\,.
\end{align}
Therefore,
\begin{align}
A^q(Q^2)=\mathcal{I}_{1}^q(Q^2)\,.
\end{align}

\begin{align}
&\langle P+q, \uparrow|T_q^{++}|P, \downarrow\rangle + \langle P+q, \downarrow|T_q^{++}|P, \uparrow\rangle \nonumber\\
&=2(P^+)^2\int\frac{d^2k_{\perp}dx}{16\pi^3}  x 
\bigg[\big\{\psi_{+\frac{1}{2}}^{\uparrow *}(x,\vec{k'}_{\perp})\psi_{+\frac{1}{2}}^{\downarrow}(x,\vec{k}_{\perp})
+ \psi_{-\frac{1}{2}}^{\uparrow *}(x,\vec{k'}_{\perp})\psi_{-\frac{1}{2}}^{\downarrow}(x,\vec{k}_{\perp})\big\}\nonumber\\
&+\big\{\psi_{+\frac{1}{2}}^{\downarrow *}(x,\vec{k'}_{\perp})\psi_{+\frac{1}{2}}^{\uparrow}(x,\vec{k}_{\perp})
+ \psi_{-\frac{1}{2}}^{\downarrow *}(x,\vec{k'}_{\perp})\psi_{-\frac{1}{2}}^{\uparrow}(x,\vec{k}_{\perp})\big\}\bigg] \, \nonumber\\
&=2(P^+)^2 ~(iq^2_{\perp})~ 2 \int dx~ N_1 N_2 \frac{1}{M_n} x^{a_1+a_2}(1-x)^{b_1+b_2+2}\exp\bigg[-\frac{\log(1/x)}{\kappa^2}\frac{Q^2}{4}\bigg]\nonumber\\
&= \frac{2(P^+)^2}{M_n} ~(iq^2_{\perp})~\mathcal{I}_{2}^q(Q^2)\,,
\end{align}
Using the matrix elements Eq.(\ref{tensor}),
\begin{align}
&\langle P+q, \uparrow|T_q^{++}|P, \downarrow\rangle + \langle P+q, \downarrow|T_q^{++}|P, \uparrow\rangle =B(Q^2)\frac{2(P^+)^2}{M}(iq^2_{\perp})\,,
\end{align}
Therefore,
\begin{align}
B^q(Q^2)=\mathcal{I}_{2}^q(Q^2)\,.
\end{align}
\subsection{$T^{+-}$: up going to down plus down going to up matrix elements}
\begin{align}
&\langle P+q, \uparrow|T_q^{+-}|P, \downarrow\rangle + \langle P+q, \downarrow|T_q^{+-}|P, \uparrow\rangle\nonumber\\
&=\int\frac{d^2k_{\perp}dx}{16\pi^3}  \frac{(k_{\perp}^2+m^2)}{x}
\bigg[\big\{\psi_{+\frac{1}{2}}^{\uparrow *}(x,\vec{k'}_{\perp})\psi_{+\frac{1}{2}}^{\downarrow}(x,\vec{k}_{\perp})
+ \psi_{-\frac{1}{2}}^{\uparrow *}(x,\vec{k'}_{\perp})\psi_{-\frac{1}{2}}^{\downarrow}(x,\vec{k}_{\perp})\big\}\nonumber\\
&+\big\{\psi_{+\frac{1}{2}}^{\downarrow *}(x,\vec{k'}_{\perp})\psi_{+\frac{1}{2}}^{\uparrow}(x,\vec{k}_{\perp})
+ \psi_{-\frac{1}{2}}^{\downarrow *}(x,\vec{k'}_{\perp})\psi_{-\frac{1}{2}}^{\uparrow}(x,\vec{k}_{\perp})\big\}\bigg] \, \nonumber\\
&=2 (iq^2_{\perp})~ \int dx~ {N_1}{N_2} \frac{1}{M_n} x^{a_1+a_2-2}(1-x)^{b_1+b_2+2}\bigg[\frac{\kappa^2(1-x)^2}{\log(1/x)}+\frac{Q^2(1-x)^2}{4}+m^2\bigg]\exp\bigg[-\frac{\log(1/x)}{\kappa^2}\frac{Q^2}{4}\bigg]\nonumber\\
&= \frac{(iq^2_{\perp})}{M}~{\mathcal{I}_{4}^q(Q^2) }.
\end{align}
Here $m$ is the quark mass which is zero in the present model. Using the matrix elements Eq.(\ref{tensor}),
\eq
&&{\langle P+q, \uparrow|T_q^{+-}|P, \downarrow\rangle + \langle P+q, \downarrow|T_q^{+-}|P, \uparrow\rangle }\nonumber\\
&&{=\big[A(Q^2)(2M)-B(Q^2)\frac{Q^2}{M}+C(Q^2)\frac{4Q^2}{M}+\bar{C}(Q^2)(4M)\big](-iq^2_{\perp})} \,.
\en
Therefore,
\eq
\big[A(Q^2)(2M)-B(Q^2)\frac{Q^2}{M}+C(Q^2)\frac{4Q^2}{M}+\bar{C}(Q^2)(4M)\big]= \frac{\mathcal{I}_{4}^q(Q^2)}{M}\,.
\en
\subsection{$T^{-2}$ : up going to down plus down going to up matrix elements}
\begin{align}
&\langle P+q, \uparrow|T_q^{-2}|P, \downarrow\rangle + \langle P+q, \downarrow|T_q^{-2}|P, \uparrow\rangle \nn\\ 
&= - \frac{4}{P^+} \int \frac{d^2k_{\perp} dx}{16\pi^3} \left(\frac{k^{\perp}_2(-(k^{\perp}_1)^2+(k^{\perp}_2)^2+m^2)}{x(1-x)}\right) \Big[
\psi^{\uparrow*}_{+\frac{1}{2}}( x,k'_{\perp})\psi^{\downarrow}_{+\frac{1}{2}}( x,k_{\perp}) \nn \\ &
+ ~\psi^{\uparrow*}_{-\frac{1}{2}}( x,k'_{\perp})\psi^{\downarrow}_{-\frac{1}{2}}( x,k_{\perp}) + ~\psi^{\downarrow*}_{+\frac{1}{2}}( x,k'_{\perp})\psi^{\uparrow}_{+\frac{1}{2}}( x,k_{\perp}) + ~ \psi^{\downarrow*}_{-\frac{1}{2}}( x,k'_{\perp})\psi^{\uparrow}_{-\frac{1}{2}}( x,k_{\perp}) 
\Big]\nn \\
&=  - \frac{1}{P^+} i (q^2_{\perp})^2 \int dx {N_1}{N_2} \frac{1}{M} x^{a_1+a_2-2} (1-x)^{b_1+b_2+2} \left( \frac{1}{x}\right)^{-\frac{Q^2}{4\kappa^2}} \nn \\
&\times\Bigg(
\frac{4(x-1)^2 \kappa^2 }{\mathrm{log}(1/x)} + \left(Q^2(x-1)^2-4m^2 \right)
\Bigg)\nn \\
&= - \frac{1}{P^+}\frac{ i  (q^2_{\perp})^2}{2M}  ~\mathcal{I}_{3}^{q}(Q^2)
\end{align}
 Using the matrix elements Eq.(\ref{tensor}),
\be
\langle P+q, \uparrow|T_q^{-2}|P, \downarrow\rangle + \langle P+q, \downarrow|T_q^{-2}|P, \uparrow\rangle = - \frac{1}{P^+}\frac{i  (q^2_{\perp})^2  }{2M} \Big[2 \mathrm{A(Q^2) M^2 }- \left(\mathrm{B(Q^2) }-4 \mathrm{C(Q^2)}\right) Q^2  \Big]
\label{eqm2lhs}\ee
Therefore,
\be
2 \mathrm{A(Q^2) M^2 }- \left(\mathrm{B(Q^2) }-4 \mathrm{C(Q^2)}\right) Q^2 = \mathcal{I}_{3}^q(Q^2).
\ee

\subsection{$T^{-1}$ : up going to down plus down going to up matrix elements}

\begin{align}
&\langle P+q, \uparrow|T_q^{-1}|P, \downarrow\rangle + \langle P+q, \downarrow|T_q^{-1}|P, \uparrow\rangle \nn\\
 &= - \frac{4}{P^+} \int \frac{d^2k_{\perp} dx}{16\pi^3} \left(\frac{k^{\perp}_1(-(k^{\perp}_1)^2+(k^{\perp}_2)^2+m^2)}{x(1-x)}\right) \Big[
\psi^{\uparrow*}_{+\frac{1}{2}}( x,k'_{\perp})\psi^{\downarrow}_{+\frac{1}{2}}( x,k_{\perp}) \nn \\ & 
+ ~\psi^{\uparrow*}_{-\frac{1}{2}}( x,k'_{\perp})\psi^{\downarrow}_{-\frac{1}{2}}( x,k_{\perp}) + ~\psi^{\downarrow*}_{+\frac{1}{2}}( x,k'_{\perp})\psi^{\uparrow}_{+\frac{1}{2}}( x,k_{\perp}) + ~ \psi^{\downarrow*}_{-\frac{1}{2}}( x,k'_{\perp})\psi^{\uparrow}_{-\frac{1}{2}}( x,k_{\perp}) 
\Big]\nn \\
&=  - \frac{1}{P^+} i q^1_{\perp} q^2_{\perp} \int dx {N_1}{N_2} \frac{1}{M} x^{a_1+a_2-2} (1-x)^{b_1+b_2+2} \left( \frac{1}{x}\right)^{-\frac{Q^2}{4\kappa^2}} \nn \\
&\times\Bigg(
\frac{4(x-1)^2 \kappa^2 }{\mathrm{log}(1/x)} + \left(Q^2(x-1)^2-4m^2 \right)
\Bigg)\nn \\
&= - \frac{1}{P^+} \frac{i q^1_{\perp} q^2_{\perp}}{2M} ~\mathcal{I}_{5}^q(Q^2).
\end{align}
Using the matrix elements Eq.(\ref{tensor}),
\be
\langle P+q, \uparrow|T_q^{-1}|P, \downarrow\rangle + \langle P+q, \downarrow|T_q^{-1}|P, \uparrow\rangle = - \frac{1}{P^+}\frac{i q^1_{\perp} q^2_{\perp} }{2M} \Big[2 \mathrm{A(Q^2) M^2 }- \left(\mathrm{B(Q^2) }-4 \mathrm{C(Q^2)}\right) Q^2  \Big]
\label{eqm1lhs}\ee
Therefore,
\be
2 \mathrm{A(Q^2) M_n^2 }- \left(\mathrm{B(Q^2) }-4 \mathrm{C(Q^2)}\right)Q^2 = \mathcal{I}_{5}^q(Q^2)=\mathcal{I}_{3}^q(Q^2).
\label{eqm1lhs}\ee

\end{document}